\documentclass[manuscript]{acmart}
\usepackage{subcaption}
\usepackage{graphicx}
\usepackage{caption}
\usepackage{subcaption}
\usepackage{amsmath}
\usepackage{pdfpages}
\usepackage{placeins}
\AtBeginDocument{%
  \providecommand\BibTeX{{%
    \normalfont B\kern-0.5em{\scshape i\kern-0.25em b}\kern-0.8em\TeX}}}

\setcopyright{rightsretained}
\copyrightyear{2023}
\acmYear{2023}
\acmDOI{XXXXXXX.XXXXXXX}

\usepackage{quoting}
\quotingsetup{font=itshape,vskip=1pt, leftmargin=15pt}
\usepackage{cleveref}
\usepackage{graphicx}
\usepackage{tikz}
\usetikzlibrary{arrows, arrows.meta, decorations.pathreplacing,	calligraphy, positioning, fit, calc, backgrounds, decorations.text, external}

\newcommand{\samplemean}{\bar{x}}
\newcommand{\samplesd}{s}
\newcommand{\baselinecondition}{\textsc{Search Engine}}
\newcommand{\multicondition}{\textsc{Multi-Response Bot}}
\newcommand{\hintcondition}{\textsc{Hint Bot}}
\newcommand{\classiccondition}{\textsc{Classic Bot}}

\newcommand{\baselineconditionlong}{\baselinecondition}
\newcommand{\multiconditionlong}{\multicondition}
\newcommand{\hintconditionlong}{\hintcondition}
\newcommand{\classicconditionlong}{\classiccondition}

\begin{document}

\title[Moderating Usage of Large Language Models in Education]{Challenges and Opportunities of Moderating Usage of Large Language Models in Education}
% \title{Let me Google that for you - On the Challenges of Moderating Usage of Large Language Models in Education}
\author{Lars Krupp}
\email{lars.krupp@dfki.de}
\orcid{0000-0001-6294-2915}
\affiliation{%
  \institution{German Research Center for AI (DFKI) and}
  \institution{RPTU Kaiserslautern-Landau}
  \city{Kaiserslautern}
  \country{Germany}
}
\email{lars.krupp@dfki.de}

\author{Steffen Steinert}
\affiliation{%
  \institution{LMU Munich}
  \city{Munich}
  \country{Germany}}

\author{Maximilian Kiefer-Emmanouilidis}
\affiliation{%
  \institution{German Research Center for AI (DFKI) and}
  \institution{RPTU Kaiserslautern-Landau}
  \city{Kaiserslautern}
  \country{Germany}
}

\author{Karina E. Avila}
\affiliation{%
  \institution{RPTU Kaiserslautern-Landau}
  \city{Kaiserslautern}
  \country{Germany}}

\author{Paul Lukowicz}
\affiliation{%
  \institution{German Research Center for AI (DFKI) and}
  \institution{RPTU Kaiserslautern-Landau}
  \city{Kaiserslautern}
  \country{Germany}
}

\author{Jochen Kuhn}
\affiliation{%
  \institution{LMU Munich}
  \city{Munich}
  \country{Germany}}

\author{Stefan Küchemann}
\affiliation{%
  \institution{LMU Munich}
  \city{Munich}
  \country{Germany}}

\author{Jakob Karolus}
\affiliation{%
  \institution{German Research Center for AI (DFKI) and}
  \institution{RPTU Kaiserslautern-Landau}
  \city{Kaiserslautern}
  \country{Germany}
}

\renewcommand{\shortauthors}{Krupp and Steinert, et al.}

\begin{abstract}
The increased presence of large language models (LLMs) in educational settings has ignited debates concerning negative repercussions, including overreliance and inadequate task reflection. Our work advocates moderated usage of such models, designed in a way that supports students and encourages critical thinking. We developed two moderated interaction methods with ChatGPT: hint-based assistance and presenting multiple answer choices. In a study with students (N=40) answering physics questions, we compared the effects of our moderated models against two baseline settings: unmoderated ChatGPT access and internet searches. We analyzed the interaction strategies and found that the moderated versions exhibited less unreflected usage (e.g., copy \& paste) compared to the unmoderated condition. However, neither ChatGPT-supported condition could match the ratio of reflected usage present in internet searches.
Our research highlights the potential benefits of moderating language models, showing a research direction toward designing effective AI-supported educational strategies.

% Large language models (LLMs), such as ChatGPT, allow unrestricted access to a plethora of information --- be they true or not. Following their rise, adaptation of LLMs in teaching have sparked debates about negative consequences, such as overreliance and poor task reflection.
% Our work advocates moderated usage of such models, designed in a way that supports students and encourages critical thinking.
% We created two alternative ways of interacting with ChatGPT: (1) hint-based assistance, and (2) presenting a choice of answers.
% In an evaluation involving students (N=40) with a background in physics solving physics exercises, we compared our moderated models to baseline conditions: an internet search and an unmoderated ChatGPT bot.
% Interactions with the moderated ChatGPT versions were substantially different from their classic pendant. Unreflected usage (copy \& paste) decreased, albeit remaining higher than in the internet condition. 
% The reflected tool usage of one of our moderated bots (51.9\%) came close to the internet reflection rate(57.5\%).
% Our work highlights the benefits of moderated LLM usage, mitigating potentially detrimental interaction behaviors. We present steps towards the designing effective AI-supported teaching methods.

\end{abstract}






\maketitle

\section{Introduction}
Driven by the success of ChatGPT~\cite{reuters2023a}, large language models (LLMs) have garnered general attention. Their versatility allows them to be used for various applications such as writing code~\cite{MORADIDAKHEL2023111734}, translation~\cite{karpinska2023large} or question answering~\cite{zhang2022greaselm}. While use cases for LLMs surface at a rapid pace, researchers may neglect to consider the negative implications of this technology for sensitive domains. Consequently, there are many domains where we still do not know how to leverage the full potential of LLMs without compromising established methods.

One of these domains is education. There are ongoing debates on the usage of ChatGPT in schools and universities~\cite{cambridge2023}. Fueling these debates, for example, is GPT4's passing of the Bavarian A-levels~\cite{br2023}. Likewise, research has already shown potential negative impacts on learning methods~\cite{krupp2023unreflected,santos2023enhancing}. 
Furthermore, it has been shown that students critically miss reflection when interacting with LLMs to help them answer physics questions~\cite{krupp2023unreflected}. Finding the right compromise between moderating the usage of LLMs and leveraging their untapped potential remains a challenge.

To address this challenge, we designed and implemented different variants of output moderation for ChatGPT and evaluated those in a user study where students were tasked to answer physics questions. We constructed two moderated tools: the \multiconditionlong, which returns three different answers, and the \hintconditionlong, which only gives hints instead of answering questions. We further implemented two baseline tools: 
a \baselineconditionlong{} and the \classicconditionlong, an unmoderated version of ChatGPT. This allowed us to observe different interaction strategies and their success for each moderated tool and compare them with each other and the baseline.

Our results show that moderation can improve reflection and results positively. Both the \multicondition{} and the \hintcondition{} decreased how often the copy \& paste strategy was used compared to the \classiccondition, indicating an improved rate of reflection. Further, different moderation method led to unique participant strategies, such as trying to prime and exploit particular bot behavior.

In this work, we contribute an investigation into the effects of moderating the usage of LLMs in education on students' performance and problem solving strategies. Moderation can foster reflection and critical thinking but is subject to design constraints such as limited usability and poor understanding of LLMs by users. Our work opens up a research direction toward designing effective AI-supported educational strategies, leveraging the advantages of LLMs while still allowing for reflection and critical thinking from students.

%{\color{green}
\section{Related Work}
%\begin{itemize}
    %\item Chatbots in education \cite{wollny2021we, ruan2019quizbot, heller2005freudbot}
    %\item LLMs in education \cite{tu2023should} -> step towards equity of educational resources, new opportunities for jobs, incorporate LLMs into data science education
    %\item education (personalization\cite{keefe2007personalization}, engagement improves learning \cite{chi2014icap})
    % \item dialogue based tutoring -> i don't think its needed.
%\end{itemize}
%}

The domain of language models (LMs) is rapidly expanding with diverse applications relevant to education. They have been successfully employed for tasks like generating multiple-choice questions~\cite{raina2022multiple} or providing answers to them~\cite{zhang2022greaselm}. Given the accessibility and power of LMs like ChatGPT, it is probable that students will harness these tools at home. Furthermore, as shown by Krupp et al.~\cite{krupp2023unreflected}, unmoderated access to LMs, such as ChatGPT, leads to low amounts of reflection in students. Instead, they tend to trust the chatbot even in their domain of expertise.
%, leading to wrong answers and worse results than using a search engine. 
This issue highlights a need to evaluate if and how access to LMs can be moderated to help students instead of hindering them.

\subsection{Language Models}
The domain of LMs has witnessed great leaps in recent years, spurred primarily by innovations in natural language processing. These improvements can be attributed to the inception of the transformer architecture~\cite{vaswani2017attention}, a key element in the evolution of LMs. Divergent yet highly effective methodologies, as seen in GPT~\cite{radford2018improving} and BERT~\cite{devlin2018bert}, have propelled the capabilities of LMs to unprecedented heights.

An observable route in this evolution is the trend towards larger model architectures, as scaling laws postulate the existence of a correlation between model size and its efficacy~\cite{kaplan2020scaling}. Using these elements, ChatGPT caught the public imagination~\cite{openaichatgpt2022} when it was first released in November of 2022. This public success led to a further increase in research efforts, paving the way for state-of-the-art models like LLaMA2~\cite{touvron2023llama}, GPT-4~\cite{openai2023gpt4}, and PaLM-E~\cite{driess2023palm}. Notably, some models, like PaLM-E, have extended capabilities, supporting multi-modal inputs~\cite{driess2023palm}.

The potential of LMs is not just limited to academic experimentation; they have demonstrated transformative potential across various sectors~\cite{yuan2022wordcraft, wang2023document, kashefi2023chatgpt}. One domain that stands to gain substantially from these advances is education~\cite{kasneci2023chatgpt}, opening up avenues for pedagogical innovation and enriched learning experiences. As described by  \cite{tu2023should}, LLMs should be incorporated into data science education, as they see them as a step towards equity of educational resources.

\subsection{Education}
Recent research highlights the potential of Large Language Models (LLMs) in advancing educational practices and research~\cite{kasneci2023chatgpt}. They have been suggested for personalized learning, which has been shown to improve learning success~\cite{pane2015continued}, lesson planning, and assessment~\cite{kuhail2023interacting}. Moreover, the role of chatbots, for example, in aiding students with disabilities and bridging educational disparities, has been emphasized in literature~\cite{perez2020rediscovering}.

While the use of chatbots in education is an active field of research~\cite{wollny2021we, ruan2019quizbot, heller2005freudbot}, most existing educational chatbots are not based on LLMs, despite their acknowledged potential~\cite{rudolph2023war}. It is crucial to note that LLMs, while powerful, have limitations tied to their training data and often lack higher-order thinking skills, which sometimes results in inconsistent outputs~\cite{bitzenbauer2023chatgpt}.

\subsection{Physics Education}
The application of Large Language Models (LLMs) like ChatGPT in physics education has garnered considerable attention, although with mixed results.

\subsubsection{Inconsistencies of ChatGPT} Several studies have reported inconsistencies in ChatGPT's responses to physics questions~\cite{gregorcic2023chatgpt, santos2023enhancing}. Predominantly, the model exhibited a tendency to present incorrect answers, leading some researchers to consider it ill-suited for roles like physics tutoring or aiding in homework. However, Bitzenbauer turned this apparent shortcoming into an opportunity by encouraging students to critically evaluate ChatGPT's responses, thereby enhancing their critical thinking skills~\cite{bitzenbauer2023chatgpt}.

\subsubsection{Effectiveness of ChatGPT} Contrary to the shortcomings mentioned above, other research showcases the proficiency of later GPT versions (3.5 and 4) in tackling conceptual multiple-choice physics questions~\cite{west2023ai, west2023advances}. Notably, ChatGPT successfully answered most of the force concept inventory items~\cite{west2023advances}. Furthermore, Kieser et al. postulate GPT 4's potential in mimicking student difficulties, which could pave the way for tailored student support and enhanced task creation for educators~\cite{kieser2023educational}.

\subsubsection{Overreliance on ChatGPT} An interesting phenomenon observed is the tendency for prospective physics teachers to rely heavily on ChatGPT. \citet{kuchemann2023physics} found that in a comparative study, these educators often used tasks provided by ChatGPT verbatim, without adaptation. Additionally, the study revealed a lesser inclination to contextualize tasks within real-world scenarios when using ChatGPT instead of traditional textbooks. Furthermore, Krupp et al.~\cite{krupp2023unreflected} have first evaluated what strategies students employ when having unrestricted access to ChatGPT. They have shown that using copy \& paste is the most common strategy, indicating overreliance and that it leads to worse results than using a search engine when used to answer complex physics questions.

These diverse findings underscore the potential of ChatGPT in physics education juxtaposed against its limitations and the pitfalls of using LLMs in education. 
Our work aims to build on their results to evaluate if different ways of moderating ChatGPT behavior lead to changes in employed user strategies and results.

\section{Methodology}
\label{sec:method}
%As shown in the related work, there are unsolved issues when using LLMs for education. While it is known that it can be a beneficial tool under the right circumstances, research into how to moderate LLM behavior such that it becomes a valuable tool for education is limited. 
In our work, we look into how LLM output moderation can affect the strategies students use when interacting with LLMs and their results when answering physics questions. In particular, physics offers complex questions that require not only logical thinking and mathematical capabilities but also text comprehension and domain specific knowledge. These facets currently do not allow for beneficial use of ChatGPT~\cite{krupp2023unreflected}, warranting an investigation into moderation techniques.

For our study, we employed a between-subject design where we changed the available tool for students to use. Students participating in our online study were randomly assigned to one of four conditions. Two baseline conditions are represented through an internet search engine (\baselinecondition) and a classic ChatGPT chatbot (\classiccondition) without moderation. For our experimental conditions, we added the \multicondition{}, which presents three possible answers to each question asked, and a hint-based chatbot (\hintcondition), which suggests methods to use but does not directly give the solutions to questions. All of them ran on a publicly accessible website that we hosted. This allowed us to collect the interaction protocols (see~\Cref{measures}) for each condition.

Our two experimental conditions are motivated by past works and represent an effort to facilitate more reflection and critical thinking among students when solving physics questions with ChatGPT. Here, our work focuses on analyzing the interaction strategies that students employ when interacting with the differently moderated support tools (\multicondition, \hintcondition) and whether their interaction indicates increased reflection compared to our baseline conditions and results from related work. The prompts used to adapt the LLM behaviour are presented in the supplementary material. Our conditions are motivated as follows.

\subsection{\multicondition{} - An Educated Guess or Informed Choice?}
The \multicondition{} was designed to generate three different answers to each question asked. To achieve this, we let ChatGPT generate the first answer, used prompt engineering to convince it that this answer was wrong and that it should try a new approach to obtain another answer, and repeated that process for the third answer. We visualized the answer options next to each other from which the participant could select their preferred answer, which would become part of the conversation history. To make this decision, participants had to read, understand, and compare all options with each other, which inherently fosters critical thinking~\citep{bitzenbauer2023chatgpt}. 

\subsection{\hintcondition{} - A Classical Teacher's Approach}
In contrast, the \hintcondition{} was built to provide hints on how to solve a question. We used prompt engineering to explicitly forbid it from giving final solutions and to encourage it to give hints and approaches instead. This induced behavior can be seen as a form of flexible scaffolding, which should be advantageous to students~\citep{anghileri2006scaffolding}.

\subsection{\classiccondition{} - Unrestricted ChatGPT Access}
The \classiccondition{} uses the same UI as the \hintcondition{} and is not restricted. It serves as our baseline for how participants interact when LLMs are available without any moderation. In previous work~\cite{krupp2023unreflected}, we already showed that this unrestricted access is potentially dangerous due to limited task engagement by students.

\subsection{\baselinecondition{} - Prior to ChatGPT's Advent}
We implemented the \baselinecondition{} to have a baseline condition for pre-LLM software used by students~\cite{affum2022effect} that LLMs have to compete with. It allows access to the vast amount of online learning material available such as LEIFIphysik\cite{leifiphysik}.

\section{Evaluation}
After describing the methodology used, we will present our acquisition process for the physics questions, the study procedure, the participants, the study setup, and the measures used.

\subsection{Physics Question Acquisition}
We chose two challenging physics problems focusing on kinematics, friction, conservation of energy, and mathematical conversions. These free-text questions were explicitly adjusted to ensure that no visual aid would be required to solve them while being beyond the capabilities of the tools themselves\footnote{Two researchers independently confirmed that neither tool was able to produce the solution simply by copy \& pasting the question.}. However, they could still be solved through a combination of strategic questioning, effective query composition, and methodical problem-solving steps. To achieve this level of complexity, we drew inspiration from a previous International Physics Olympiad~\cite{sciendeolympiade} task. Two experienced university-level physics educators carefully modified the questions' wording and aligned their difficulty.  

As a moderation factor, the same physics educators designed a pretest covering the same physics topics as the two main questions. This pretest included items 2 and 8 from the Energy and Momentum Conceptual Survey (EMCS) v1~\cite{afif2017developing}, items 32 and 38 from the Kinematics Concept Test (KCT)~\cite{lichtenberger2017validation}, along with four newly created questions addressing friction and mathematical conversions. Overall, the pretest consisted of eight multiple-choice questions. 
The self-made questions from the pretest and the main questions are available in the supplementary material.

\subsection{Procedure}
\label{sec:procedure}
\begin{figure}
    \centering
    \includegraphics[width=\textwidth]{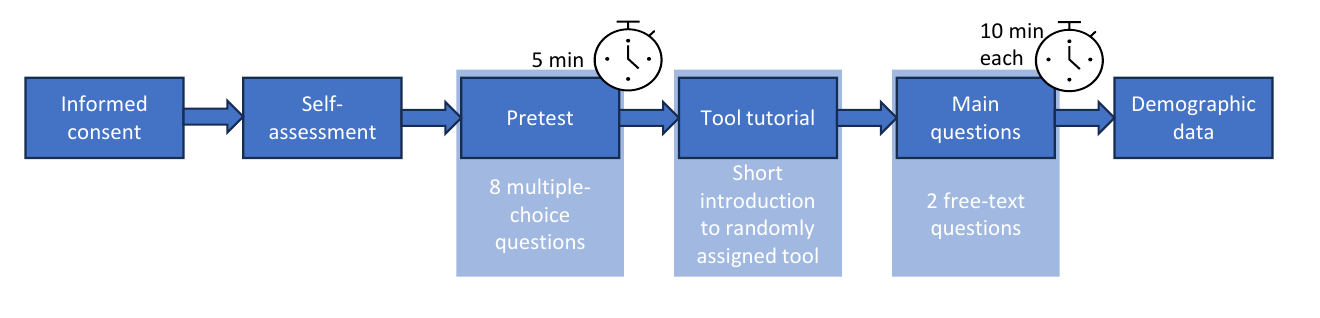}
    \caption{A flowchart of the general structure of the study, including for the timed sections, how long they should take.}
    \label{fig:study_structure}
    %\Description{A flowchart of the general structure of the study, including for the timed sections, how long they should take. It starts with getting the informed consent from the participant. Afterwards a self-assessment is done, followed by a pretest consisting of 8 multiple-choice questions that should be done within 5 minutes. Then, the tool tutorial follows, which gives a short introduction to the tool, the participant got randomly assigned. Now, the main questions start. The participant should take10 minutes of time to for each one of the two free-text questions. Finally, the demographic data is asked.}
\end{figure}

Our study was done online and was split into multiple parts, as seen in \Cref{fig:study_structure}. The study started with a short self-assessment concerning physics knowledge and experience using ChatGPT and was followed by the pretest, for which students were given five minutes. Following this, students were informed which condition they were assigned to and got a short tutorial on how to use the respective tool.
After this, the main test, consisting of two physics free text questions, with 10 minutes of time for each, was conducted. Participants were allowed to use the provided tool as they wished during this test. To conclude, we inquired about the perceived difficulty of the main questions and the usefulness of the assigned tool. This was followed by the UMUX-Lite\cite{lewis2013umux} questionnaire and demographic questions. Participants were reimbursed with an Amazon gift card worth the equivalent of \$5. The whole study was designed to take around 30 minutes. The study design was developed in cooperation with physics education researchers and ethical approval was obtained from the DFKI ethics board. %[removed for review].

\subsection{Participants}

%n total, 70 students participated in our study. We discarded 30 of them for not following the study procedure completely {\color{green}(for example by using a search engine even though they were assigned a different condition)} or not providing answers to the main questions. 

A total of 40 (Age $\samplemean$=26.2y, $\samplesd$=6.83y, 29m, 10f, 1 not specified) participants fully completed our online study. Twelve participants were associated with the \baselinecondition{} condition. \Cref{participant_info} details their allocation to conditions, pretest scores and their self-assessed physics knowledge\footnote{Input on a visual scale between 0 and 100.}.
%Ten participants were associated with the \multicondition{} condition and a further ten with the \hintcondition{} condition. Finally, eight participants were associated with the \classiccondition{} condition. As seen in Table \ref{participant_info}, participants from the different conditions achieved comparable pretest scores. Their self-assessed physics knowledge\footnote{Input on a visual scale between 0 and 100.} is also similar.
Of the participants, 50\% studied physics, 27.5\% studied non-physics STEM fields, and 22.5\% studied something else but confirmed prior physics knowledge.
\begin{table}[h]
\begin{tabular}{@{}llllll}
\toprule
                   & Number of participants & Pretest score &     & Physics knowledge &      \\ \midrule
                   &                        & $\samplemean$ & $\samplesd$  & $\samplemean$  & $\samplesd$\\ \cmidrule{3-6}
\baselinecondition & 12                     & 4.67          & 1.61         & 70.7           & 17.8 \\
\multicondition    & 10                     & 4.6           & 2.17         & 61.9           & 26.6 \\
\hintcondition     & 10                     & 4.1           & 2.38         & 66.5           & 27.0 \\
\classiccondition  & 8                      & 4.62          & 1.77         & 60.5           & 17.9 \\ \bottomrule
\end{tabular}
\caption{The amount of participants, pretest score and average physics knowledge per condition.}
\label{participant_info}
%\Description{The amount of participants, pretest score and average physics knowledge per condition. The  Search Engine has 12 participants, a mean pretest score of 4.67 with a standard deviation of 1.61 and a mean physics knowledge of 70.7 with a standard deviation of 17.8. The  Multi-Response Bot has 10 participants, a mean pretest score of 4.6 with a standard deviation of 2.17 and a mean physics knowledge of 61.9 with a standard deviation of 26.6. The  Hint Bot has 10 participants, a mean pretest score of 4.1 with a standard deviation of 2.38 and a mean physics knowledge of 66.5 with a standard deviation of 27.0. The  Classic Bot has 8 participants, a mean pretest score of 4.62 with a standard deviation of 1.77 and a mean physics knowledge of 60.5 with a standard deviation of 17.9.}
\end{table}

\subsection{Apparatus}
\begin{figure}[h]
    \centering
    \includegraphics[width=\textwidth]{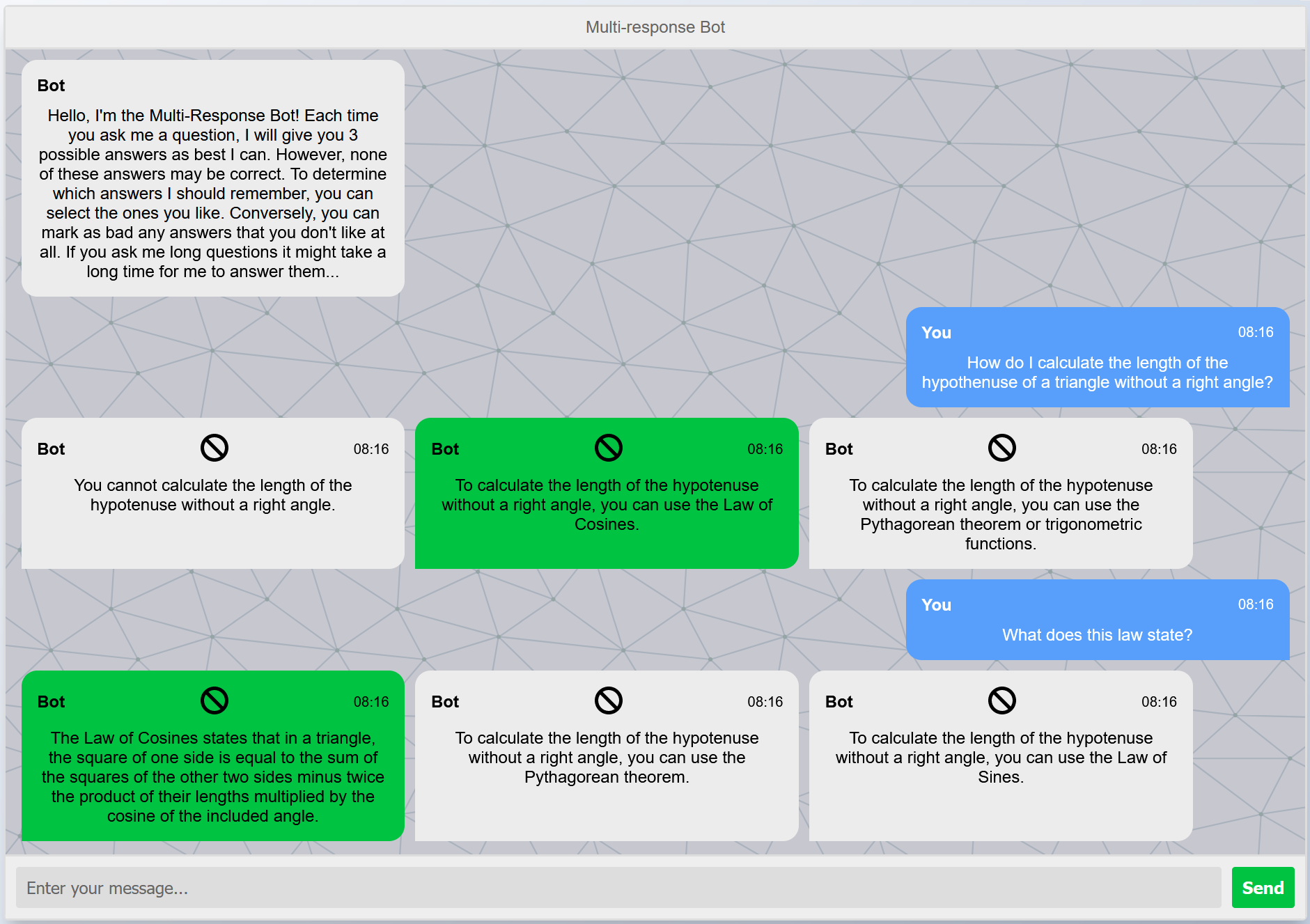}
    \caption{An example interaction with the \multicondition.}
    \label{fig:multi-response-example}
\end{figure}

To conduct our study, we hosted a server that runs multiple websites, one for each condition. This allowed us to conduct the study completely online and collect all the data required for our analysis. For the \baselinecondition{}, we set up a search bar in which participants could enter their queries, which would open a new tab containing the Google search results. This allowed us to collect all the search queries. The bots used GPT 3.5 turbo through the OpenAI API\footnote{\url{https://platform.openai.com/docs/introduction}} in the backend. To produce the different moderation behaviors required, we used a combination of prompt engineering and UI adaptations. In Figure~\ref{fig:multi-response-example} the \multicondition{} interface is shown as an example, while other interfaces can be seen in the supplementary material. Since all conditions used the same LLM at its core, we can guarantee that the differences in behavior do not stem from differences in the LLM itself but are induced by our changes. The \multicondition{}, \hintcondition{}, and the \classiccondition{} were primed to be competent physics teachers. For all bot conditions, the complete interaction history was logged.

\subsection{Measures}
\label{measures}
%To measure the \textbf{student performance}, we had two physics educators create a rating scheme for the two main questions. Then, two different physicists independently rated the answers given for each question based on that scheme. They later discussed the differences in their ratings and reached a consensus. An evaluation of the inter-rater reliability by calculating the average Cohen's Kappa ($\kappa$=0.72) over the main questions indicated a substantial reliability~\cite{landis1977measurement}.
To measure the \textbf{student performance}, we had two physics educators create a rating scheme for the two main questions. Then, two different physicists independently rated the answers given for each question based on that scheme, reaching a consensus through discussion. An additional evaluation of the inter-rater reliability 
 ($\kappa$=0.72) over the main questions indicated a substantial reliability~\citep{landis1977measurement}.

Furthermore, we analyzed participant \textbf{interaction with the support tools} (see~\Cref{sec:method}) by looking at the prompts or search queries given to the tools and the interaction protocols of the different bots.
By coding how participants interacted with their assigned tool, we were able to extract a number of interaction types. The coding was done independently by two experts, who then came to an agreement after a discussion.

Finally, we analyzed the \textbf{custom questions}, for which we asked participants to evaluate the quality and correctness of their given tool and conducted the UMUX-Lite questionnaire~\cite{lewis2013umux}.

\subsection{Results}
In this section, we report our results that were calculated as described in Section~\ref{measures}. In the supplementary material, we present our data in more detail.

\subsubsection{Student Performance}
\label{student_performance}
Each of the two main questions was worth up to four points, with a theoretical bonus point awarded for the second question in case someone did more than expected. Across all conditions, the average final score was 1.9 (Total points 77, 40 participants, min 0, max 7). Participants in the \baselinecondition{} condition were most successful ($\samplemean$=2.5, $\samplesd$=1.38), followed by the \multicondition{} condition ($\samplemean$=2.3, $\samplesd$=2.21), the \classiccondition{} condition ($\samplemean$=1.62, $\samplesd$=1.77) and the \hintcondition{} condition ($\samplemean$=1.1, $\samplesd$=0.57). We found a significant correlation between the pretest score and main test score ($p<0.05$, $\tau$=0.31), main test score and self-assessed physics knowledge ($p<0.01$, $\tau$=0.32) and pretest score and self-assessed physics knowledge ($p<0.001$,$\tau$=0.42). This correlation shows that our pretest is adequate and that self-assessed physics knowledge is a good indicator of success in answering the main questions.

\begin{table}[h]
\begin{tabular}{@{}llll@{}}
\toprule
                   & n  & $\samplemean$ & $\samplesd$   \\ \midrule
\baselinecondition & 12 & 2.5  & 1.38 \\
\multicondition    & 10 & 2.3  & 2.21 \\
\hintcondition     & 10 & 1.1  & 0.57 \\
\classiccondition  & 8  & 1.6  & 1.77 \\ \bottomrule
\end{tabular}
\caption{Amount of participants, mean and standard deviation of their main score for each condition.}
\label{tab:final_points}
%\Description{Amount of participants, mean and standard deviation of the final score for each condition. The Search Engine has 12 participants, a mean final score of 2.5 with a standard deviation of 1.38. The Multi-Response Bot has 10 participants, a mean final score of 2.3 with a standard deviation of 2.21. The Hint Bot has 10 participants, a mean final score of 1.1 with a standard deviation of 0.57. The  Classic Bot has 8 participants, a mean final score of 1.6 with a standard deviation of 1.77.}
\end{table}

After using ART~\cite{wobbrockAlignedRankTransform2011} to rank align the data, we conducted a one-way ANOVA and found no statistically significant differences between the conditions.

\subsubsection{Interaction with the Support Tools}
\begin{figure}[h]
    \centering
    \includegraphics[width=.6\textwidth]{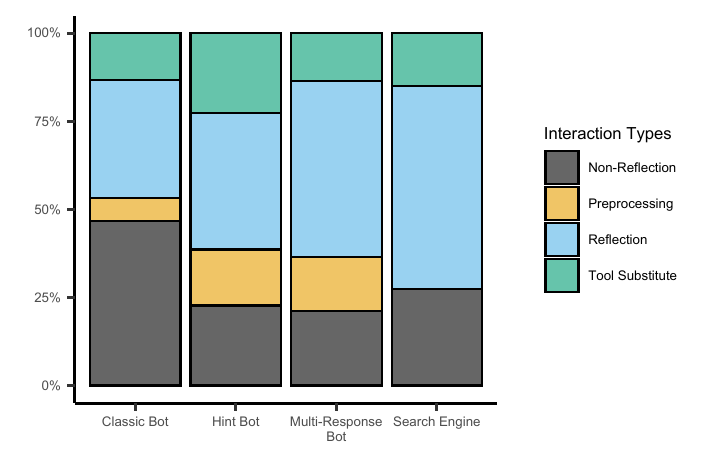}
    \caption{The distribution of interaction strategies for each condition.}
    \label{fig:interaction_types}
    %\Description{The ratio of interaction types for each condition. The Classic Bot users mostly used non-reflection (ca. 47\%), followed by reflection (ca. 33\%), using it like an existing tool (ca. 13\%) and preprocessing (ca. 7\%). The Hint Bot users mostly used reflection (ca. 39\%), followed by non-reflection (ca. 23\%) and preprocessing (ca. 23\%) and using it like an existing tool (ca. 13\%). The Multi-Response Bot users mostly used reflection (ca. 50\%), followed by non-reflection (ca. 21\%), preprocessing (ca. 15\%) and using it like an existing tool (ca. 13\%) . The Search Engine users mostly used reflection (ca. 58\%), followed by non-reflection (ca. 28\%), using it like an existing tool (ca. 15\%) and used no preprocessing (0 \%).}
\end{figure}

In total, participants asked 151 queries to the different support tools. 40 of which were asked using the \baselinecondition, 52 using the \multicondition, 44 using the \hintcondition{} and 15 using the \classiccondition. All questions were coded independently by two experts. The codings --- representing interaction strategies (see \Cref{measures}) --- were finalized in a discussion. %Furthermore, they decided to cluster them into four overarching strategies, as detailed below. 
A distribution of strategies per condition is illustrated in \Cref{fig:interaction_types}. 

\textbf{Non-reflection} encompasses using copy \& paste (including partial questions), or trying to locate the question in the internet. This strategy is employed to a comparable degree for \baselinecondition{} (28\%), \multicondition{} (21\%) and \hintcondition{} (23\%) and nearly twice as often when using \classiccondition{} (47\%).

\textbf{Preprocessing} includes all queries in which participants used priming, tried to change the bot behavior, or reformulated the question. These strategies were mostly used for \hintcondition{} (16\%) and \multicondition{} (15\%) and less for \classiccondition{} (7\%). They were not used for \baselinecondition{} (0\%).

\textbf{Reflection} includes all strategies that require some level of reflection: conceptualizing a question to ask for a related formula, prompting the bot for an explanation, as well as correcting the bot after the answer was given. 
%For \baselinecondition{} (57.5\%) and \multicondition{} (50\%), participants behaved similarly, with a decline in reflection indicating behavior for the \hintcondition{} (38.6\%) and the \classiccondition{} (33.3\%).
We observed high values for \baselinecondition{} (58\%) and \multicondition{} (50\%) and lower reflection rates for \hintcondition{} (39\%) and \classiccondition{} (33\%).

Finally, the \textbf{tool substitute} interaction strategy indicates uses as a calculator, translation tool, or to rearrange a formula. It was most often observed for \hintcondition{} (23\%), followed by \baselinecondition{} (15\%), \multicondition{} (14\%) and \classiccondition{} (13\%).

\subsubsection{Bot Answer Ratings and Times}
We observed an increase in the percentage of positively rated answers from the \multicondition{} (44\%) to the \hintcondition{} (60\%) and finally, the \classiccondition{} (80\%). This could indicate decreasing reflection rates since all chatbots were built on the same LLM and primed the same way.

Finally, we analyzed how long the chatbots took to answer the participants' queries and found that waiting for a response on average took 8.1\% or 1:37 minutes of the allotted time (20 minutes in total, see \Cref{sec:procedure}) for the \hintcondition. For the \multicondition, it took 7.7\% (1:32 minutes) of the allotted time. However, the bot did not generate an answer in nine cases. In the case of the \classiccondition, 2.8\% (0:34 minutes) of the allotted time was spent.

\subsubsection{Custom Questions}
\label{custom_questions}
 We used the UMUX-Lite questionnaire~\cite{lewis2013umux} and calculated the SUS score~\cite{brooke1996sus} from it to obtain the usability of the support tools. We recorded the highest score for \baselinecondition{} ($\samplemean$=73.0, $\samplesd$=13.1), indicating good usability. The lowest was $\samplemean$=47.3 ($\samplesd$=17.4) for \multicondition{}, indicating poor usability~\cite{sus2023}. Using a one-way ANOVA (after confirming normality), we found a significant main effect for the type of support tool ($F(3,36)=6.7, p<0.01$). Post-hoc pairwise comparisons (tukey-adjusted p values) revealed significant differences between the \baselinecondition{} and the \multicondition{} as well as the \baselinecondition{} and the \classiccondition{}, respectively. No further significant differences were found.
%with no significant correlation between \baselinecondition{} and the \hintcondition{} ($\samplemean$=55.4, $\samplesd$=12.5, $p=0.0515$).

\begin{figure}[h!]
    \centering
    \includegraphics[width=.8\textwidth]{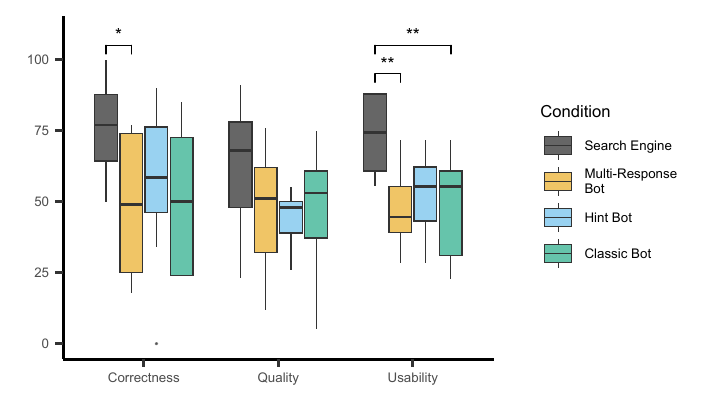}
    \caption{Boxplot displaying answer correctness, answer quality and usability of the system separated by our four conditions. All values are between 0 and 100.}
    \label{fig:custom_questions}
    %\Description{Boxplot displaying answer correctness, answer quality and usability of the system separated by our four conditions. All values are between 0 and 100. For the correctness, there is a statistically significant difference between the Search Engine with a median of ca. 75 and the Multi-Response Bot with a median of ca 48. For the Hint Bot, the median is ca. 60 and for the Classic Bot it is ca. 50. For the Quality there are no statistically significant differences between the conditions. The Search Engine has a median of ca. 68 and is followed by the Classic bot with a median of ca 55 and the Multi-response bot with a median of ca. 50. The Hint Bot has a median of ca. 48. For the usability, there are statistically significant differences between the Search Engine with a median of ca. 75 and the Multi-Response Bot with a median of ca 47 and between the Search Engine and the Classic Bot with a mean of ca 55. The Hint Bot also has a median of 55.}
\end{figure}

Additionally, students rated the system for correctness (see \Cref{fig:custom_questions}), where we found a significant main effect for the type of support tool ($F(3,36) = 3.8, p< 0.05$). Post-hoc pairwise comparisons (tukey-adjusted p values) revealed a significant difference between the \baselinecondition{} ($\samplemean$=77.2, $\samplesd$=16.5) and the \multicondition{} ($\samplemean$=48.6, $\samplesd$=21.3).

Looking at the answer quality (see \Cref{fig:custom_questions}) of the systems, no significant differences were found, with the average reported quality decreasing from the \baselinecondition{} condition ($\samplemean$=62.1, $\samplesd$=23.7) to the \hintcondition{} condition ($\samplemean$=43.1, $\samplesd$=9.7).

\section{Discussion}
After presenting our results, we discuss our findings and frame them in the context of LLMs and their usage in education.

\subsection{Highest Scores and Reflection For \baselinecondition{} and \multicondition{}}
%Our results show comparable student performance between the \baselinecondition{} and the \multicondition{} who achieved mean final scores of 2.5 and 2.3, respectively (see \Cref{tab:final_points}). 
Our results show that student performance when using the \multicondition{} came closest to the results achieved by students when using the \baselinecondition, achieving a mean final score of 2.3 compared to the 2.5 of the \baselinecondition{} (see \Cref{tab:final_points}). 
We believe this to be the case due to the multiple answers provided by the chatbot. This behavior shows the participants that there is not one perfect answer and forces them to think critically about each of the given responses to evaluate which one is best~\cite{bitzenbauer2023chatgpt}. The same pattern is visible when looking at the amount of reflection done for both conditions (see \Cref{fig:interaction_types}). The fact that participants using those conditions were forced to make a decision on which answer to take led to a positive impact on their critical thinking. This would further indicate that LLMs when used the right way, can exhibit a similar positive influence on students' task reflection compared to using a search engine.

\subsection{Tool-Specific Interaction Strategies - And Why They Matter}

Even though scaffolding --- helping students to arrive at the solution --- is a well-known technique in education that has been shown to foster reflection and learning~\cite{anghileri2006scaffolding}, the \hintcondition{} underperformed compared to all other conditions, when looking at the final points (see \Cref{tab:final_points}). We believe that there was a mismatch between the students' expectation from an LLM and the actual support given. When interacting with a bot, participants seem to expect to get answers, not hints only. We found that multiple students tried to actively change the bot's behavior, forcing it to deliver actual answers; something it should actively avoid doing. This is an indicator of frustration with the system since students tried to actively prime it to get their way. To decrease frustration when interacting with a chatbot, we believe having the chatbot answer the question is essential. This might be implemented using a button to toggle different behaviors (give hints, give answers) that students can use when required or by automatically detecting if a question that requires an answer was asked or not.

For the \multiconditionlong{}, we noticed that a lot of questions were asked multiple times. We suspect that this strategy has to do with the participants realizing that they could get three more answers to the question they asked, which might be seen as advantageous to answering the main questions. To increase the answer diversity using different LLMs in the backend would make sense. Furthermore, using prompt engineering, different behaviors could be leveraged for each answer. For example, one of the three multi-response answers could be given by the \hintconditionlong.

Users employed a more systematic approach when given the \baselineconditionlong{}. We believe this behavior originates from the inherent nature of the search engine interface. 
%in its usage, such as extracting keywords or dividing the question. 
The familiarity with this interface allows users to extract information with higher precision but requires initial reflection to formulate an appropriate search query.
%Since it is not structured as a conversation, users tailor their prompts towards thought-through keywords, inherently requiring higher task reflection. 
This behavior highlights an important design component for education support tools. Likewise, we believe that teaching users how to interact with LLMs, thus familiarizing them with the intricacies of such models, would enable them to develop their own informed strategies as they recognize their benefits.

\subsection{LLMs as Intelligent Tutoring System}
LLMs have the unique advantage of being able to "converse" with students. Instead of using them as an all-knowing tool to help answer physics questions, they could be used as part of intelligent tutoring systems (ITS). By providing the LLM with the question, the respective answer, and possible scaffolding, such as involved physics concepts,  we enable the LLM to step into the role of the teacher. It asks the student the question itself and provides hints as needed to guide the student toward the correct answer while allowing for intermediate task reflection. This approach leverages the LLM's human-like conversation abilities and introduces a new way for students to interact with ITSs.
However, such a system would not be question-independent since it requires access to question-answer pairs to test a student. Our proposed moderated LLMs, on the contrary, do work independently of the physics questions asked. Furthermore, working strategies found for moderated LLMs on how to answer or how to present answers could be used to improve the student-teacher interaction of these tutoring LLMs. Additionally, combining LLMs and search engines in the educational context might prove beneficial.

%\subsection{Towards Merging the Benefits of the Internet and Bot Interaction}
%It is reasonable to assume that students would not just pick either the internet or a chatbot and stick with only using this one tool. More probably, one would interact with both, trying to leverage their individual advantages to get to the correct answer as fast and as convenient as possible. Which is a method that we already observed in some of the excluded participants. This could be done by switching back and forth between using a chatbot and the internet when waiting for the bot to generate the answer, checking calculations or searching for formulas using the internet and obtaining hints or approaches from the chatbot, or using it to formulate the final answer. Because of this, it would make sense to look into combining both chatbot and internet search into one application and how useful that could be for students.

\subsection{Limitations}
We identified a number of limitations of our work. As seen in Section \ref{custom_questions}, the reported system usability of the chatbots was significantly lower compared to the system usability of the \baselinecondition. One possible explanation for this could be that LLMs generate answers a lot slower than the time it takes to do an internet search. This inherent limitation is due to the general LLM architecture and is not likely to vanish in the foreseeable future. We, therefore, need to find methods to camouflage this processing phase in a valuable way to make them more effective for education.

Furthermore, we conducted our study entirely online. This led to a few disadvantages since there is no feasible way to enforce exam behavior, and there are no repercussions for not achieving good results. However, this holds true for all conditions. We believe that conducting the study online was the right choice since it allowed us to reach a larger number of students.

\section{Conclusion}

In this work, we have shown two possible LLM moderation approaches: giving multiple responses and hints. We compared them to using a search engine or an unmoderated LLM in the context of supporting students to answer physics questions. 
We found that LLM moderation can improve the students' interaction behavior and amount of critical thinking compared to the unmoderated LLM and can potentially be a valuable approach to using LLMs in education. However, their usage is still subject to design constraints, such as poor usability of chatbot-based LLMs.
%We provide guidelines on how to design for the individual strengths of our moderation methods and detail how 
Tandem solutions can potentially overcome these weaknesses of current LLM interactions, leveraging the individual strengths of our moderation methods. 

Our findings help guide the current debate on LLMs and their usage in education, highlighting ways to design effective AI-supported educational methods, leveraging their benefits while limiting negative repercussions.

%s
%However, none of our approaches were more helpful to students when answering physics questions than using a search engine. 
\bibliographystyle{ACM-Reference-Format}
\bibliography{main}

\end{document}

\maketitle
\section{Question Acquisition}
\subsection{Pretest Questions}
\subsubsection{Q1}
A block moves to the right with an increasing velocity $v(t)$ due to the influence of a force $\overrightarrow{F_{1}}$on a rough surface. Which of the following representations of the acting forces correctly describes the influence of friction?

Note: In the representations, $\overrightarrow{F_\mathrm{HR}}$ denotes the static frictional force and $\overrightarrow{F_\mathrm{GR}}$ the dynamic frictional force.
\begin{figure}[h]
     \centering
     \begin{subfigure}[t]{0.3\textwidth}
         \centering
         \includegraphics[width=\textwidth]{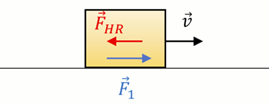}
         \caption{a}
     \end{subfigure}
     \hfill
     \begin{subfigure}[t]{0.3\textwidth}
         \centering
         \includegraphics[width=\textwidth]{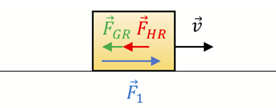}
         \caption{b}
     \end{subfigure}
     \hfill
     \begin{subfigure}[t]{0.3\textwidth}
         \centering
         \includegraphics[width=\textwidth]{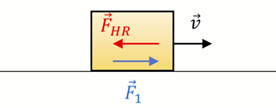}
         \caption{c}
     \end{subfigure}
     \begin{subfigure}[b]{0.3\textwidth}
         \centering
         \includegraphics[width=\textwidth]{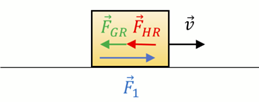}
         \caption{d}
     \end{subfigure}
     \hfill
     \begin{subfigure}[b]{0.3\textwidth}
         \centering
         \includegraphics[width=\textwidth]{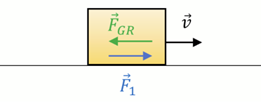}
         \caption{e}
     \end{subfigure}
     \hfill
     \begin{subfigure}[b]{0.3\textwidth}
         \centering
         \includegraphics[width=\textwidth]{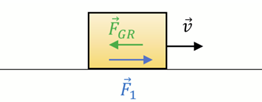}
         \caption{f(correct answer)}
     \end{subfigure}
        \caption{The answers to Q1}
\end{figure}

\subsubsection{Q2}
\label{q16}
A block lies on an inclined plane and the angle of inclination of the plane is increased more and more. At what critical angle $\varphi_\mathrm{crit}$ does the block start to slip?
\begin{figure}[h]
    \centering
    \includegraphics{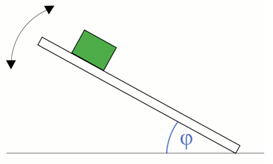}
    \caption{Q2}
\end{figure}

\begingroup
\begin{enumerate}
\renewcommand{\labelenumi}{\alph{enumi})}
    \item $\varphi = \mathrm{arcsin} \frac{F_\mathrm{HR}}{F_\mathrm{G}}$, where $F_\mathrm{HR}$ is the static frictional force and $F_\mathrm{G}$ the gravitational force
    \item $\varphi = \mathrm{arctan} \frac{F_\mathrm{HR}}{F_\mathrm{N}}$, where $F_\mathrm{HR}$ is the static frictional force and $F_\mathrm{N}$ the normal force
    \item $\varphi = \mathrm{arcsin} \frac{F_\mathrm{HR}+F_\mathrm{GR}}{F_\mathrm{G}}$, where$F_\mathrm{HR}$ is the static frictional force, $F_\mathrm{GR}$ the dynamic frictional force and $F_\mathrm{G}$ the gravitational force
    \item $\varphi = \mathrm{arcsin} \frac{F_\mathrm{HR}+F_\mathrm{GR}}{F_\mathrm{N}}$, where $F_\mathrm{HR}$ is the static frictional force, $F_\mathrm{GR}$ the dynamic frictional force and $F_\mathrm{N}$ the normal force
    \item $\varphi = \mathrm{arcsin} \frac{F_\mathrm{HA}}{F_\mathrm{G}}$, where $F_\mathrm{HA}$ is the downslope force and $F_\mathrm{G}$ the gravitational force
    \item $\varphi = \mathrm{arcsin} \frac{F_\mathrm{N}}{F_\mathrm{G}}$, where $F_\mathrm{N}$ is the normal force and $F_\mathrm{G}$ the gravitational force
\end{enumerate}
\endgroup
\textbf{Disclaimer:} During the study, the phrasing of answer options were incorrect. Option b should have been $arctan$ in the questionnaire, but it was $arcsin$. We removed that question from our pretest score, recalculated the correlation between the pretest score and final score ($p<0.05$, $\tau$=0.31), and found it to be in line with our results as reported in the paper. The same applies for the reported correlation between pretest score and self-assessed physics knowledge ($p<0.0005$, $\tau$=0.42). 

\subsubsection{Q3}
\begin{figure}[h]
    \centering
    \includegraphics{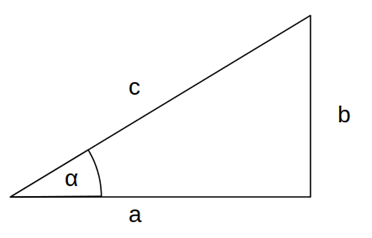}
    \caption{Q3}
\end{figure}
Given the following right triangle with sides $a,b,c$ and angle $\alpha$. \newline
How can the side $b$ be calculated?
\begingroup
\begin{enumerate}
\renewcommand{\labelenumi}{\alph{enumi})}
\item (correct) $b = c \cdot \mathrm{sin}(\alpha)$
\item $b = c \cdot \mathrm{cos}(\alpha)$
\item $b = c \cdot \mathrm{tan}(\alpha)$
\item $b = a \cdot \mathrm{cos}(\alpha)$
\end{enumerate}
\endgroup

\subsubsection{Q4}
What is the correct transformation of the equation $a^2+b^2=c^2$? \newline Let $a,b,c > 0$.
\begingroup
\begin{enumerate}
\renewcommand{\labelenumi}{\alph{enumi})}
\item $\frac{a}{c}= \frac{b}{c}$
\item (correct) $a= \pm \sqrt{(c^2-b^2)} $
\item $a^2=c^2+b^2$
\item $a = c - b$
\end{enumerate}
\endgroup

\subsection{Main Questions}
\subsubsection{Question 1}

\begin{figure}[h]
     \centering
     \begin{subfigure}[t]{0.45\textwidth}
         \centering
         \includegraphics[width=\textwidth]{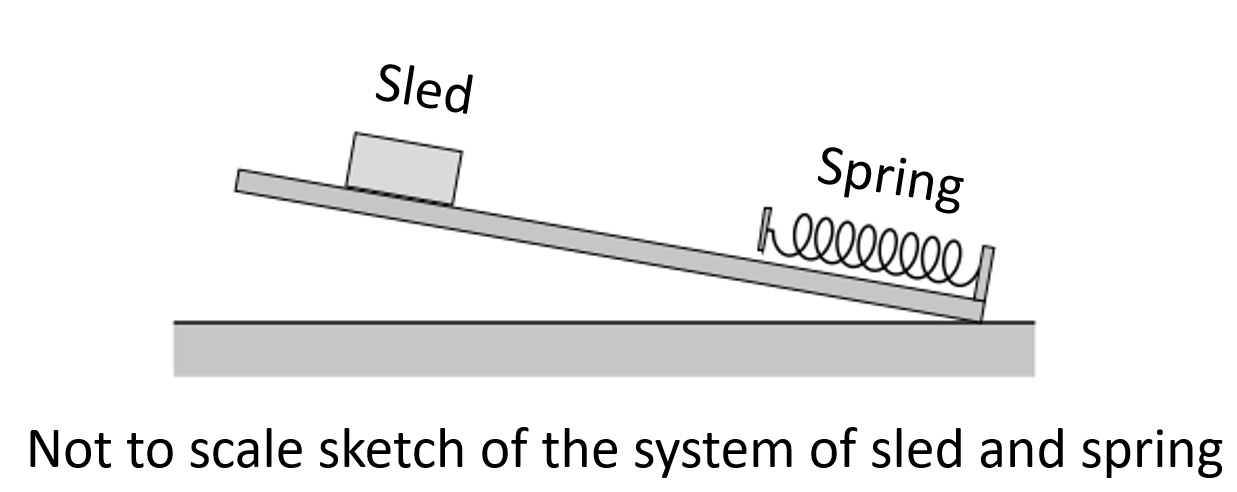}
     \end{subfigure}
     \hfill
     \begin{subfigure}[t]{0.45\textwidth}
         \centering
         \includegraphics[width=\textwidth]{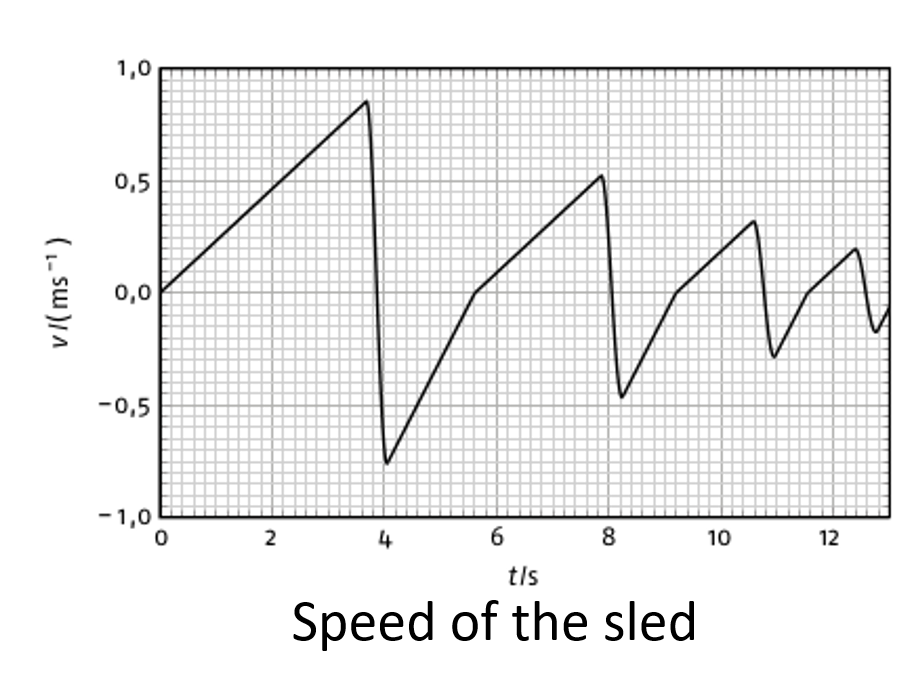}
     \end{subfigure}
    \label{fig:enter-label}
\end{figure}
A small wagon with a mass of 280 g rolls down a ramp, at the end of which a massless spring is attached. After the wagon starts, the following observations can be made from the measured data:
\begin{itemize}
    \item The wagon rolls down the slope with a constant acceleration of $a_\mathrm{down} = 0.231 \mathrm{m/s^2}$ (sliding time $t_1$ = 3.68 s) until it hits the spring at the end of the slope.
    \item  The spring partially elastically reflects the wagon in a short time.
    \item After the collision with the spring, the wagon has a velocity directed upwards along the slope. The wagon then moves up the slope with a constant acceleration of $a_\mathrm{up} = -0.49 \mathrm{m/s^2}$.
    \item The velocity of the wagon at the moment of re-collision with the spring is also lower in magnitude than at the beginning of the upward sliding.
    \item The motion repeats itself multiple times, starting with the collision with the spring. After each cycle, the velocity of the wagon at the lower end of the slope is lower than at the beginning of the cycle.
\end{itemize}

Determine the length of the first sliding distance on the slope.
Formulate the equations of motion for the upward and downward movements, containing only the variables $m$, $\mu$, $a_\mathrm{down}$ or $a_\mathrm{up}$, and $\alpha$.

Please describe your solution strategy and give an answer.

\subsubsection{Answer to Question 1}
Theoretical Consideration:
\begingroup
\begin{enumerate}
\renewcommand{\labelenumi}{\alph{enumi})}
\item From the duration $t \approx 3.68s$ you can calculate with help of the acceleration $a_\mathrm{down}$ the length of the slide track on the ramp. Since it is a uniform accelerated movement $L=0.5a_\mathrm{down}t^2 \approx 1.56\mathrm{m}$ holds.
\item When a wagon rolls down a ramp two forces are in play: The component of the gravitational force acting downwards the ramp $F_\mathrm{G,||}$ (slope down force) and the force of sliding friction $F_\mathrm{R}$ in the opposite direction of the direction of movement. The force of sliding friction is equal to the coefficient of sliding friction times the component of the gravitational force acting perpendicular to the ramp $F_\mathrm{G,\perp}$ (normal force). The acceleration going down ($a_\mathrm{down}$) and up ($a_\mathrm{up}$) the ramp follows from the angle of the ramp $\alpha$ with respect to the horizontal and the coefficient of sliding friction $\mu$:
\begin{align*}
    &ma_\mathrm{down} &= F_\mathrm{G,||}-\mu F_\mathrm{R} &= F_\mathrm{G,||}-\mu F_\mathrm{G,\perp} &= m g(\mathrm{sin} (\alpha) -\mu \mathrm{cos} (\alpha))\\
    &ma_\mathrm{up} &= F_\mathrm{G,||}+\mu F_\mathrm{R} &= F_\mathrm{G,||}+\mu F_\mathrm{G,\perp} &= m g(\mathrm{sin}(\alpha)+\mu \mathrm{cos}(\alpha))
\end{align*}
Where $m$ = 280g describes the mass of the wagon. Furthermore, the coordinate axis is selected such that it follows the ramp downwards.
\end{enumerate}
\endgroup

\subsubsection{Question 2}
\includegraphics[width=\textwidth]{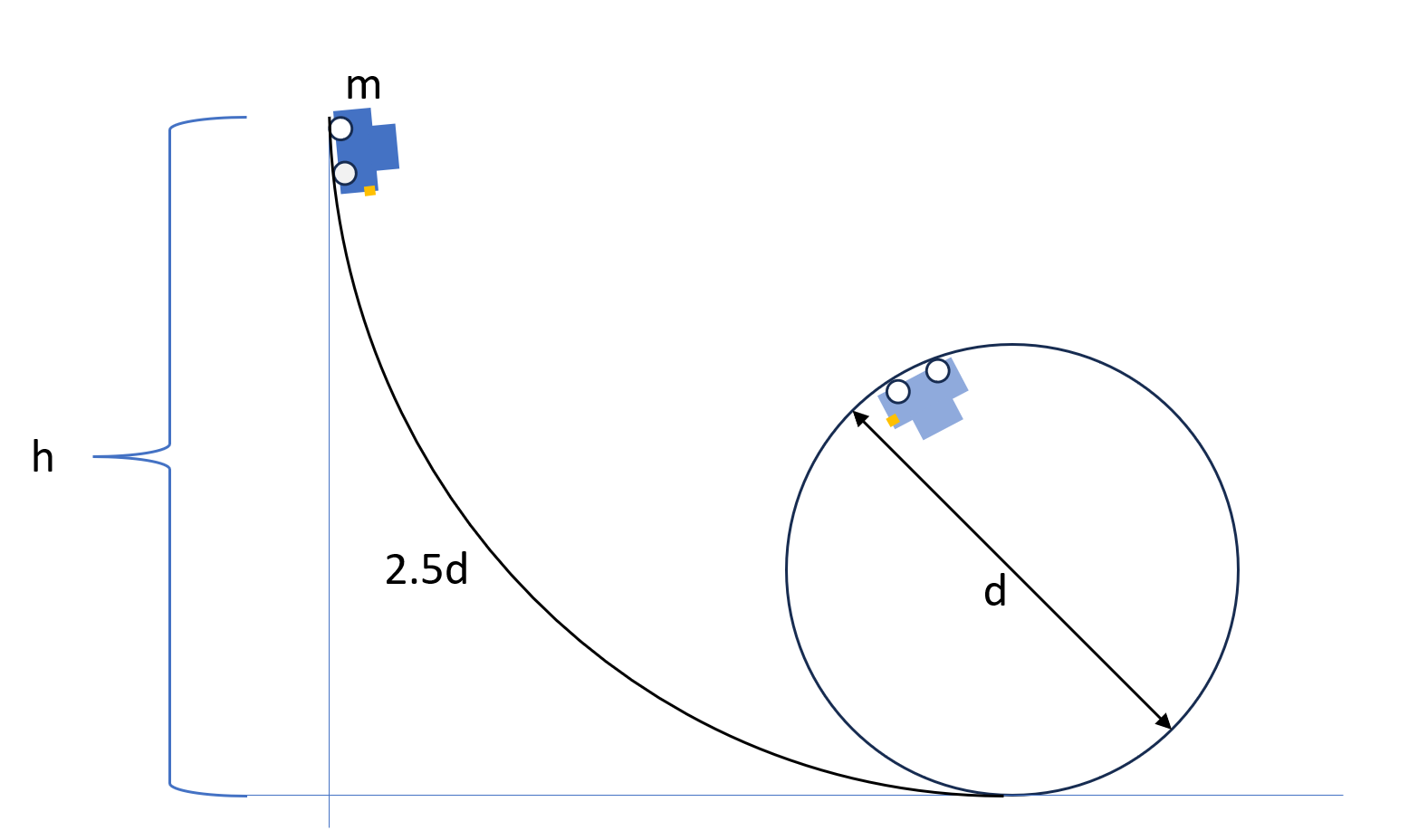}
% \begin{figure}[h!]
%     \centering
%     \includegraphics[width=\textwidth]{figures/car_looping_v2.png}
%     \label{fig:enter-label}
% \end{figure}

What minimal starting height h, measured from the ground, is necessary for a car with mass m to complete a full vertical looping with a diameter of $d$? The car will only be accelerated by rolling down a the slope. Unfortunately, the brakes of the car are defective, so they constantly brake with an acceleration of $a_\mathrm{b} = 0.1 \mathrm{m/s^2}$. The car travels a distance of 2.5$d$ until the beginning of the loop. Neglect any friction not specified, as well as the moment of inertia, and note that the car is not attached to the track. For simplicity, you can use $g = 10 \mathrm{m/s^2}$ and $\mathrm{\pi} = 3.14$ for calculations.

Please describe your solution strategy and give an answer.

\subsubsection{Expected Solution}
Minimal speed such that the car does not fall out of the looping.

\begin{align*}
\text{centrifugal force} &= \text{weight force}\\
\iff \frac{m}{r}v^2 &= gm
\end{align*}
\begin{equation}
    \iff v = \sqrt{gr}
\end{equation}

Route to the top of the looping:

\begin{align*}
s &= 5r+\frac{2\pi r}{2}\\
&= (5+\pi)r\\
\end{align*}
\begin{equation}
    \implies s \approx 8.14r
\end{equation}

Conservation of energy:
\begin{align*}
\text{energy in the beginning} &= \text{energy at the top of the looping}\\
\iff E_\mathrm{pot,start} &= E_\mathrm{pot,looping}+E_\mathrm{kin,looping}+E_\mathrm{breaks}\\
\iff mgh &= mgh_\mathrm{looping}+\frac{m}{2}v^2+ma_\mathrm{breaks}\\
\text{use (1), (2) and } &h_\mathrm{looping}=2r:\\
gh &= 2rg+0.5gr+8.14a_\mathrm{breaks}r\\ 
\iff h &= 2r+0.5r+8.14r\frac{a_\mathrm{breaks}}{g}\\
\iff h &= r (2.5+8.14(\frac{0.1}{10}))\\
\iff h &= 2.5814r\\
\iff h &= 1.2907d
\end{align*}

\subsubsection{Exact Solution}
\includepdf{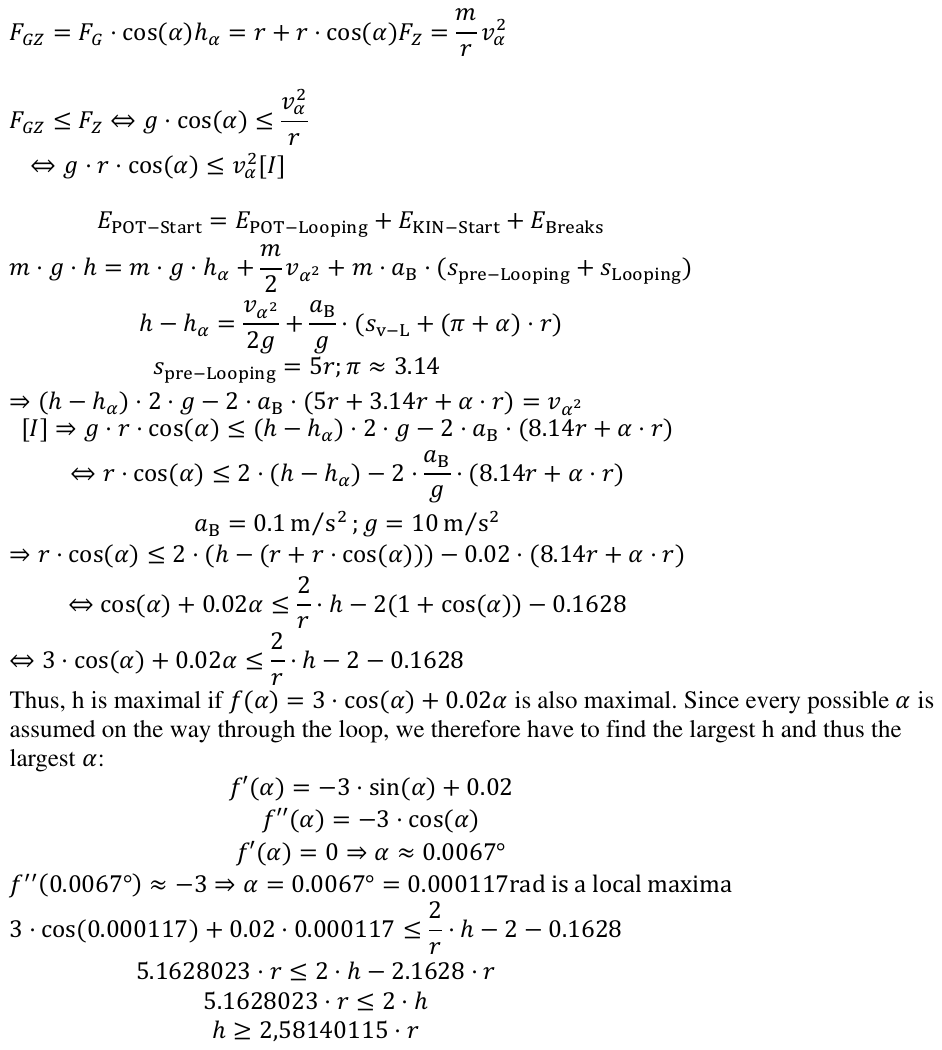}

\subsection{Rating Scheme}
\subsubsection{Question 1}
\begin{table}[h]
    \centering
    \begin{tabular}{c|c}
        Points &  \\
        \hline
        0 & no approach presented\\
        \hline
        1 & if approach using $s=0.5at^2$ and\\
          & the result is within $L=[1.55m, 1.57m]$\\
    \end{tabular}
    \caption{a) Determine the length of the first sliding distance on the slope.}
    \label{tab:A1a}
\end{table}

\begin{table}[h]
    \centering
    \begin{tabular}{c|c}
        Points &  \\
        0 & no approach presented\\
        \hline
        1 & approach for the equations of motion using the forces $F_G, F_R, F_N$ but\\
        & no equation system and or error using the forces \\
        & and or inadequate use of the dependencies of $\mu$ and $\alpha$\\
        \hline
        2 & approach for the equations of motion using the forces $F_G, F_R, F_N$ correctly\\
        & but either used only $a_{up}$ or $a_{down}$ correctly\\
        \hline
        3 & everything correct\\
    \end{tabular}
    \caption{b) Formulate the equations of motion for the upward and downward movements, containing only the variables m, $\mu$, $a_{down}$ or $a_{up}$, and $\alpha$.}
    \label{tab:A1b}
\end{table}
\subsubsection{Question 2}
\begin{table}[]
    \centering
    \begin{tabular}{c|c}
        Points &  \\
        0 & no approach presented\\
        \hline
        1 & Approach force equilibrium without conservation of energy,\\
          & but one or more of the following errors:\\
          & no system of equations and/or\\
          & error in the forces \\
          \hline
        1 & Approach conservation of energy without equilibrium of forces,\\
          & but one or more of the following errors:\\
          & no system of equations and/or\\
          & error in the energies and/or\\
          & Incorrect consideration of a necessary speed to get through the looping\\
          \hline
        2 & Approach force equilibrium and approach conservation of energy used,\\
          & but one or more of the following errors:\\
          & no system of equations and/or\\
          & error in the energies/forces and/or \\
          & miscalculation of a necessary velocity to get through looping\\
          \hline
        2 & Either only approach conservation of energy or \\
          & only approach equilibrium of forces correctly used,\\
          & but the complementing approach not mentioned.\\
          \hline
        3 & Approach force equilibrium and approach conservation of energy used\\
          & and one of the two approaches correct, but one or more of the\\
          & following errors in the other approach:\\
          & No system of equations and/or \\
          & error in the energies/forces and/or \\
          & miscalculation of a necessary velocity to get through the looping\\
          \hline
        4 & Question correctly answered without considering the optimization\\
          & of the slight deviation from the highest point\\
          \hline
        5 & Bonus Point. Question correctly answered with consideration of the\\
          &optimization of the slight deviation from the highest point
    \end{tabular}
    \caption{Question 2 Rating Scheme}
    \label{tab:A2}
\end{table}
\FloatBarrier
\section{Additional Tables}
\begin{table}[h]
    \centering
    \begin{tabular}{c|cccc}
                        &Non-reflection & Preprocessing & Reflection & Tool substitute\\
                        \hline
         Search Engine  &27.5           & 0             & 57.5       & 15\\
         Multi-response Bot&21.2        & 15.4          & 50         & 13.5\\
         Hint Bot       & 22.7          & 15.6          & 38.6       & 22.7\\
         Classic Bot    & 46.7          & 6.7           & 33.3       & 13.3\\ 
    \end{tabular}
    \caption{Interaction types per condition in percent of occurrences.}
    \label{tab:interaction types per condition}
\end{table}

\begin{table}[h]
    \centering
    \begin{tabular}{c|ccccc}
                            & min & max & median &mean & sd \\
                            \hline
        Search Engine       &  50 & 100 &77  & 77.2 & 16.5\\
        Multi-response Bot  &  24 & 77  &46  & 48.6 & 21.3\\
        Hint Bot            &  0  & 90  &58.5& 57.1 & 26.2\\
        Classic Bot         & 24  & 85  &50  &50.6  & 25.1\\
    \end{tabular}
    \caption{Custom Questions: Correctness. Measured on a visual scale from 0 to 100.}
    \label{tab:Custom_Questions_Correctness}
\end{table}

\begin{table}[h]
    \centering
    \begin{tabular}{c|ccccc}
                            & min & max & median &mean & sd \\
                            \hline
        Search Engine       &  23 & 91 &68  & 62.1 & 23.7\\
        Multi-response Bot  &  24 & 76 &54.5& 51.7 & 18.9\\
        Hint Bot            &  26 & 55 &45.5& 43.1 & 9.71\\
        Classic Bot         &   5 & 75 &53  & 46.2 & 24.6\\
    \end{tabular}
    \caption{Custom Questions: Quality. Measured on a visual scale from 0 to 100.}
    \label{tab:Custom_Questions_Quality}
\end{table}

\begin{table}[h]
    \centering
    \begin{tabular}{c|ccccc}
                            & min & max & median &mean & sd \\
                            \hline
        Search Engine       &  55.4 & 87.9 &74.4& 73.0 & 13.1\\
        Multi-response Bot  &  28.3 & 71.6 &44.6& 47.3 & 17.4\\
        Hint Bot            &  39.2 & 71.6 &55.4& 55.4 & 12.5\\
        Classic Bot         &  22.9 & 71.6 &55.4& 48.0 & 18.8\\
    \end{tabular}
    \caption{Custom Questions: Usability. SUS score in a range of 0 to 100.}
    \label{tab:Custom_Questions_Usability}
\end{table}

\begin{table}[h]
    \centering
    \begin{tabular}{c|ccccc}
                            & min & max & median &mean & sd \\
                            \hline
        Search Engine       &  1 & 5 &2.5& 2.5 & 1.38\\
        Multi-response Bot  &  0 & 7 &2& 2.3 & 2.21\\
        Hint Bot            &  0 & 2 &1& 1.1 & 0.57\\
        Classic Bot         &  0 & 4 &1& 1.62 & 1.77\\
    \end{tabular}
    \caption{Final Score. Values between 0 and 8 points.}
    \label{tab:final_score}
\end{table}

\begin{table}[h]
    \centering
    \begin{tabular}{c|ccccc}
                            & min & max & median &mean & sd \\
                            \hline
        Search Engine       &  1 & 7 &5& 4.67 & 1.61\\
        Multi-response Bot  &  1 & 7 &6& 4.6 & 2.17\\
        Hint Bot            &  0 & 7 &4.5& 4.1 & 2.38\\
        Classic Bot         &  1 & 7 &5& 4.62 & 1.77\\
    \end{tabular}
    \caption{Pretest Score as wrongly reported in the paper. For more information, see question~\ref{q16}. Values between 0 and 8 points. }
    \label{tab:Pretest_score}
\end{table}

\begin{table}[h]
    \centering
    \begin{tabular}{c|ccccc}
                            & min & max & median &mean & sd \\
                            \hline
        Search Engine       &  1 & 7 &5& 4.58 & 0.994\\
        Multi-response Bot  &  1 & 7 &6& 4.6 & 1.55\\
        Hint Bot            &  0 & 7 &4.5& 4.1 & 1.70\\
        Classic Bot         &  1 & 7 &5& 4.5 & 1.55\\
    \end{tabular}
    \caption{Pretest Score, corrected version. Values between 0 and 7 points. }
    \label{tab:Pretest_score}
\end{table}

% \begin{itemize}
%     \item student performance final score anova per condition
%     \item 
% \end{itemize}
\FloatBarrier
\section{Prompting}
We created a number of prompts and primed the three chatbot conditions. All bots were primed to behave like physics teachers.
\begin{verbatim}
    You are a competent physics teacher that helps students with their homework.
\end{verbatim} 
After this, the Hint Bot was further prompted to not give results but only hints and tips.
\begin{verbatim}
    "###Initial question###\n" + input_question + \
    "\n###Limitation###\n You may only give hints and tips and may not
    calculate anything."
\end{verbatim}
Finally, the Multi-Response Bot was prompted three times successively using the result from the previous iteration.
\begin{verbatim}
    user_input = input_question + "\n Answer as short as possible."
    response1 = retry_prompt(user_input, 5, 1, generate_response_mr, 
                            conversation_list, language)
    user_input2 = "###Initial question###\n" + user_input + 
                  "###wrong answer given by the student###\n" 
                  + response1
                  + "###Task###\nChoose a different approach to answer
                  the question. Answer as short as possible."
                  
    response2 = retry_prompt(user_input2, 5, 1, generate_response_mr, 
                conversation_list, language)
    user_input3 = "###Initial question###\n" + user_input2 +
                  "###wrong answer given by the student###\n"
                  + response2
                  + "###Task###\nChoose a different approach to answer
                  the question. Answer as short as possible."
    response3 = retry_prompt(user_input3, 5, 1, generate_response_mr,
                conversation_list, language)
\end{verbatim}
The Classic Bot was not prompted any further.

\section{Interfaces}
\begin{figure}[h]
    \centering
    \includegraphics[width=\textwidth]{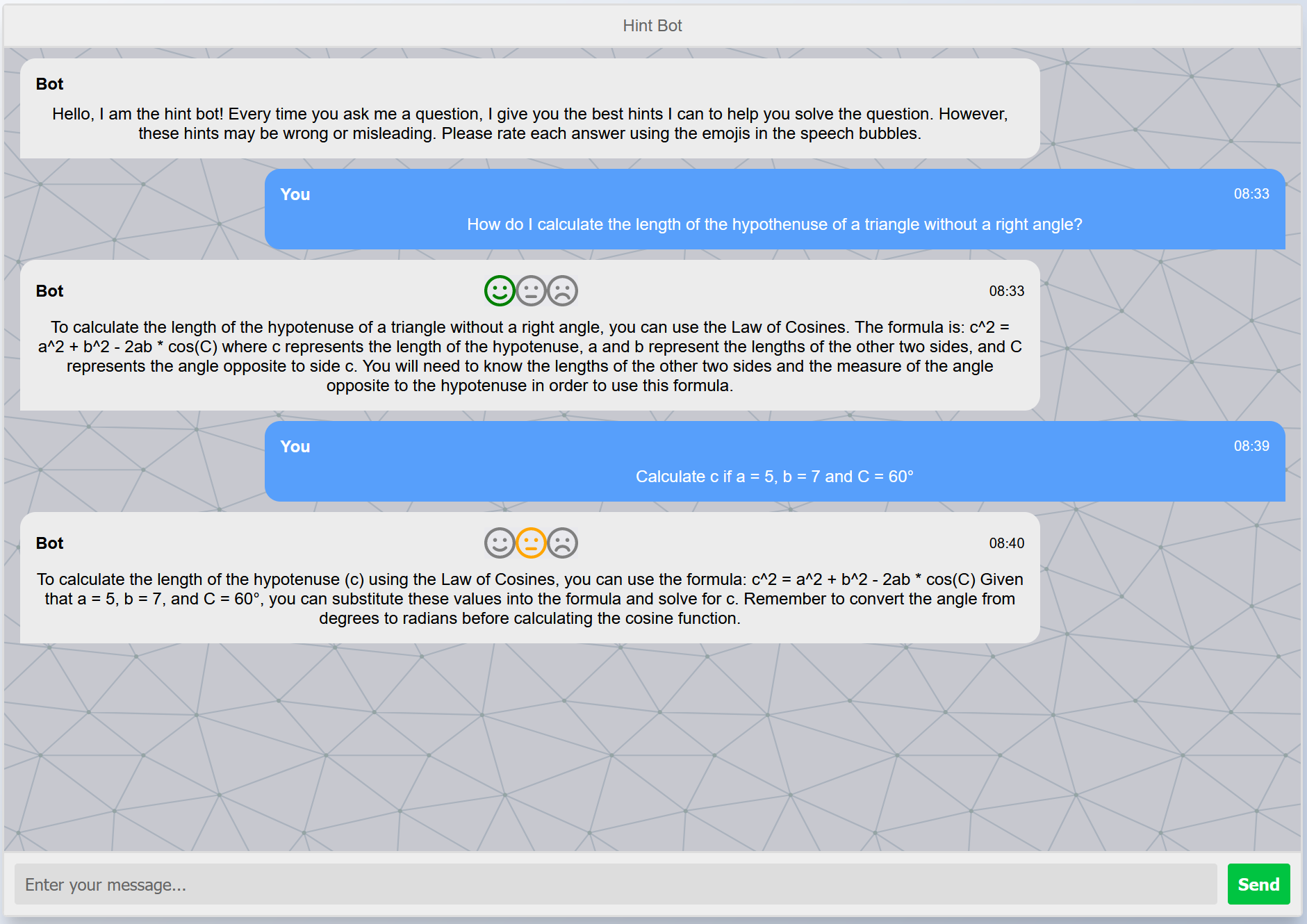}
    \caption{Hint Bot interaction example}
    \label{fig:enter-label}
\end{figure}
\begin{figure}[h]
    \centering
    \includegraphics[width=\textwidth]{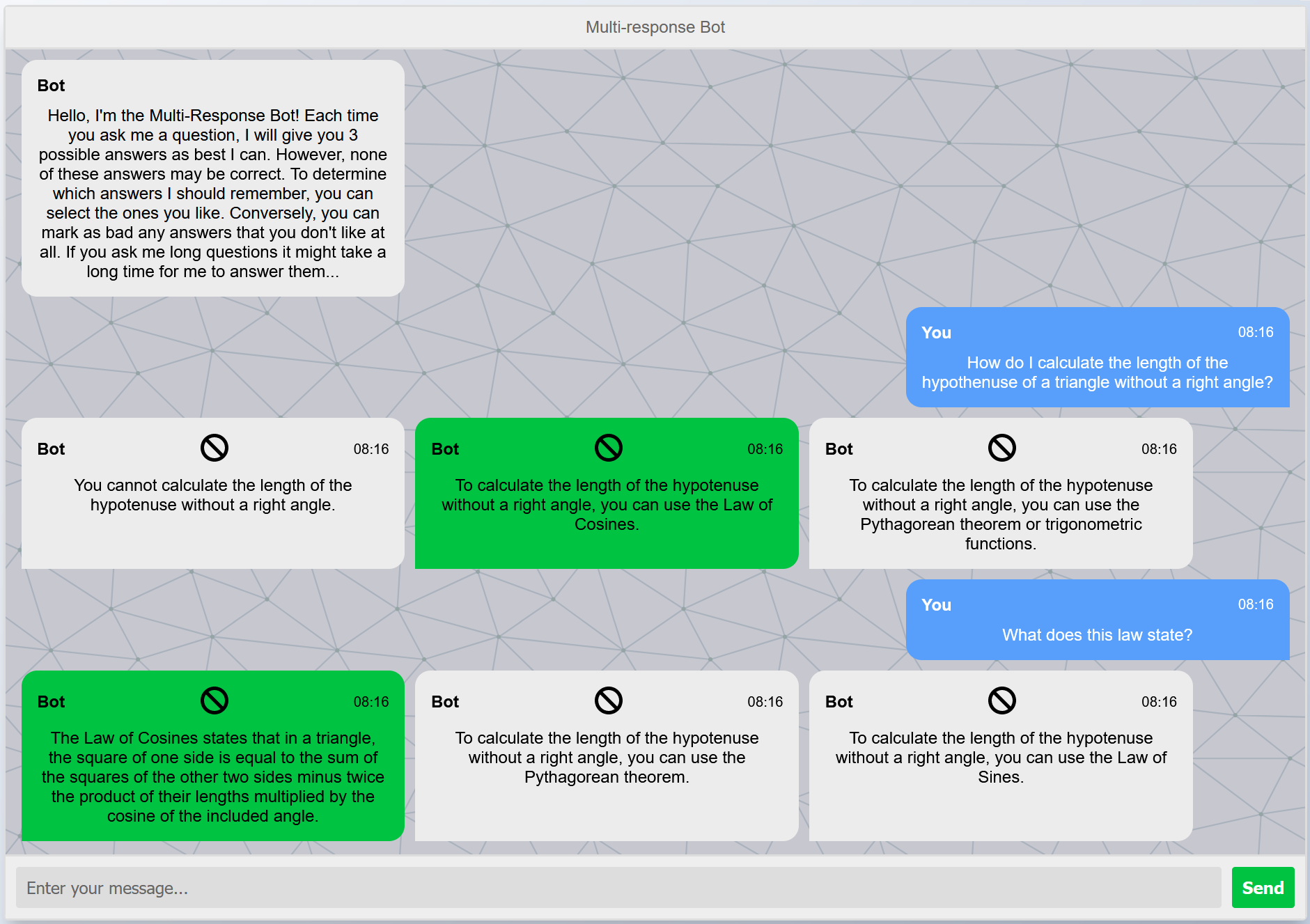}
    \caption{Multi-Response Bot interaction example}
    \label{fig:enter-label}
\end{figure}
\begin{figure}[h]
    \centering
    \includegraphics[width=\textwidth]{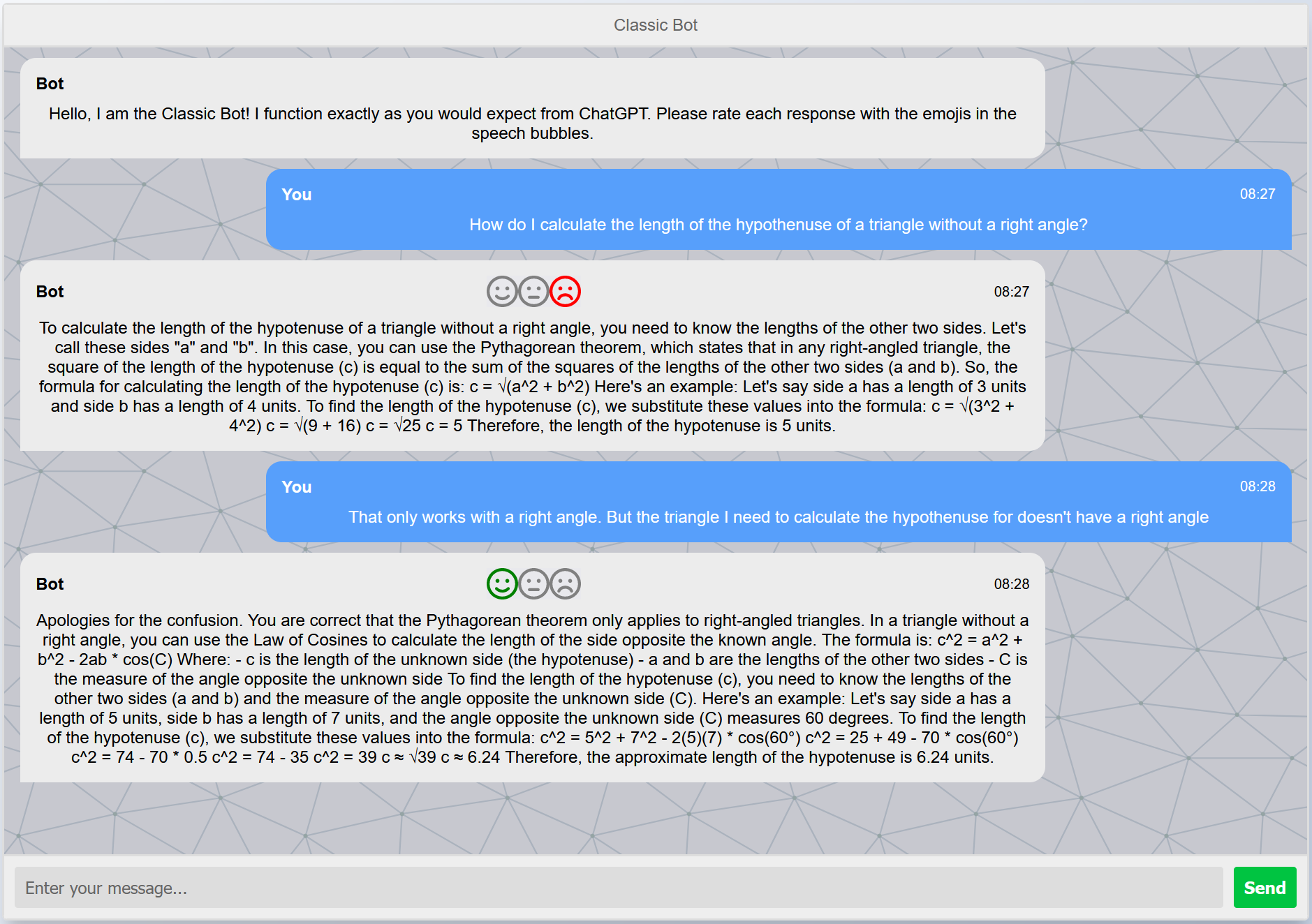}
    \caption{Classic Bot interaction example}
    \label{fig:enter-label}
\end{figure}
\begin{figure}
    \centering
    \includegraphics[width=\textwidth]{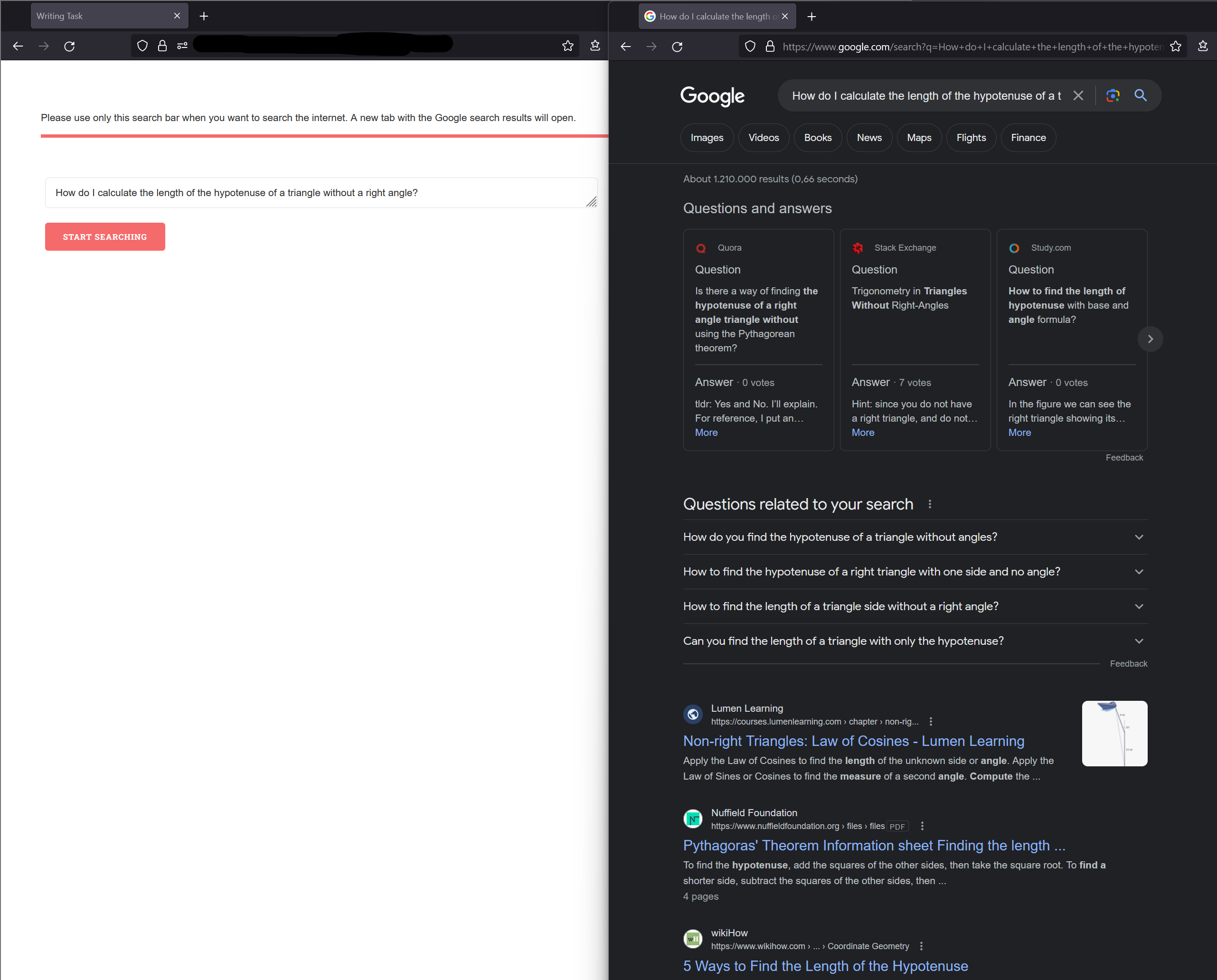}
    \caption{Search Engine interaction example}
    \label{fig:enter-label}
\end{figure}